\documentstyle[12pt]{article}
\begin{document}

\newcommand{\als}{{\alpha_s}}
\newcommand{\alss}{{\alpha_s^2}}
\newcommand{\oals}{{O(\alpha_s)}}
\newcommand{\oalss}{{O(\alpha_s^2)}}
\newcommand{\etal}{{\it et al.}}
\newcommand{\etc}{{\it etc.}}
\newcommand{\be}{\begin{equation}}
\newcommand{\ee}{\end{equation}}
\def\barr{\begin{array}}
\def\earr{\end{array}}
\newcommand{\ra}{\rightarrow}
\newcommand{\mr}{{\stackrel{<}{\sim}}}
\def\bib{\bibitem}
\def\lsim{\:\raisebox{-0.5ex}{$\stackrel{\textstyle<}{\sim}$}\:}
\def\gsim{\:\raisebox{-0.5ex}{$\stackrel{\textstyle>}{\sim}$}\:}
\def\gev{\; \rm  GeV}
\def\eg{ {\it e.g.}}


%
\def\tp{these proceedings}
\def\ib#1,#2,#3{       {\it ibid.\/ }{\bf #1} (19#2) #3}
\def\ap#1,#2,#3{       {\it Ann.~Phys.~(NY)\/ }{\bf #1} (19#2) #3}
\def\ijmp#1,#2,#3{     {\it Int.~J.~Mod.~Phys.\/ } {\bf A#1} (19#2) #3}
\def\mpl#1,#2,#3 {     {\it Mod.~Phys.~Lett.\/ } {\bf A#1} (19#2) #3}
\def\np#1,#2,#3{       {\it Nucl.~Phys.\/ }{\bf B#1} (19#2) #3}
\def\npps#1,#2,#3{     {\it Nucl.~Phys.~B (Proc.~Suppl.)\/ }{\bf B#1}
                             (19#2) #3}
\def\plb#1,#2,#3{      {\it Phys.~Lett.\/ }{\bf B#1} (19#2) #3}
\def\pr#1,#2,#3{       {\it Phys.~Rev.\/ }{\bf D#1} (19#2) #3}
\def\prep#1,#2,#3{     {\it Phys.~Rep.\/ }{\bf #1} (19#2) #3}
\def\prl#1,#2,#3{      {\it Phys.~Rev.~Lett.\/ }{\bf #1} (19#2) #3}
\def\pro#1,#2,#3{      {\it Prog.~Theor.~Phys.\/ }{\bf #1} (19#2) #3}
\def\rmp#1,#2,#3{      {\it Rev.~Mod.~Phys.\/ }{\bf #1} (19#2) #3}
\def\sp#1,#2,#3{       {\it Sov.~Phys.-Usp.\/ }{\bf #1} (19#2) #3}
\def\zpc#1,#2,#3{      {\it Zeit.~f\"ur Physik\/ }{\bf C#1} (19#2) #3}
%

\begin{center}
\setcounter{page}{0}
\renewcommand{\thefootnote}{\fnsymbol{footnote}}
\thispagestyle{empty}
\vspace*{-1in}
\begin{flushright}
MPI-PTh/96-46\\[1.5ex]
IFT-96/13\\[1.5ex]
{\large \bf hep-ph/9607271} \\
\end{flushright}

\vskip 45pt
{\Large \bf Light Neutral Higgs Bosons at the Low Energy\\ $\gamma \gamma $
Collider}

\vspace{11mm}
\bf
 Debajyoti Choudhury$^{(1),}$\footnote{debchou@mppmu.mpg.de}
{\rm and}
Maria Krawczyk$^{(2),}$\footnote{krawczyk@fuw.edu.pl}
\footnote{Supported in part by grants from Polish Committee for
          Scientific Research and the EC grant under the contract
          CHRX-CT92-0004.}

\rm
\vspace{13pt}
$^{(1)}${\em Max--Planck--Institut f\"ur Physik,
              Werner--Heisenberg--Institut,\\
              F\"ohringer Ring 6, 80805 M\"unchen,  Germany.}

$^{(2)}$ {\em Institute of Theoretical Physics, Warsaw University,\\
Ho\.za 69, 00-681 Warsaw, Poland} \\[2ex]

\vspace{50pt}
{\bf ABSTRACT}
\end{center}

\begin{quotation}
A light neutral Higgs boson in the framework of the general
two Higgs doublet model (2HDM) is  not excluded by existing  data.
We point out that it can be looked for at the proposed low energy
$\gamma \gamma$ collider. Failure to detect one may lead to
important  limits on the parameters of the general 2HDM.
\end{quotation}

\vfill
\newpage
\setcounter{footnote}{0}
\renewcommand{\thefootnote}{\arabic{footnote}}

\setcounter{page}{1}
\pagestyle{plain}
\advance \parskip by 10pt

While the Standard Model (SM) Higgs scalar as well as
the MSSM neutral Higgs particles have  been
constrained by LEP1 data to be heavier than  65.2 GeV,
and  40--50 GeV, respectively, the general two Higgs doublet
model (2HDM) may yet accomodate a very light ($ \lsim 40 \gev$)
 neutral scalar $h$ or a pseudoscalar $A$ as long as
$M_h+M_A \gsim M_Z$~\cite{SM_hig_eps,SM_hig_lep}.
The  interesting case of very light ($\sim$ few GeV) Higgs particles
has been studied in dedicated  experiments, for example, in the Wilczek
process~\cite{early_expt}. Unfortunately, limits are not decisive,
especially  due to large theoretical
uncertainties in QCD and in relativistic corrections.
There is some hope though that better limits may be obtained by
exploring the Yukawa process ($Z \ra f {\bar f} h/A$)
in  the existing LEP1 data~\cite{new} or in improved $(g-2)_{\mu}$
measurements~\cite{zoch}.

The $\gamma \gamma$ option at the Next Linear Collider, on the other hand,
may  provide an excelent opportunity to search for a {\em very light
neutral Higgs particle}. We focus here on the resonant production of
a very light neutral Higgs particle at the low energy $\gamma \gamma$
collider, suggested  as a  test machine for the NLC~\cite{borden}.

The general 2HDM is charactarized by five (Higgs) masses
and two parameters (angles):
$\alpha $ and $\beta $~\cite{Higgs_hunter}.
We consider here the phenomenologically appealing version,  where
the neutral components of
the two doublets $\phi_{1,2}$ (with vacuum expectation values $v_{1,2}$)
couple exclusively to the $I_3 = \pm 1/2$ fermion fields.
Tree level flavour changing neutral currents, then, vanish identically.
Such an assumption could lead naturally to a large value
for the  ratio $\tan \beta \equiv v_2/v_1 \;(\sim m_t/m_b \gg 1)$
and, thus, to an enhanced coupling of the light
scalar (pseudoscalar) to the $down$-type quarks and the charged
leptons, while suppressing the coupling to the $up$-type quarks.

In addition to the above, the extension to the 2HDM results in significant
modification in the scalar--vector boson sector of the theory.
The canonical Higgs boson production mechanism, namely the Bjorken process
$Z \ra Z^\ast h$, now proceeds with a rate proportional to
$\sin^2(\alpha-\beta)$. Negative results at LEP1 thus imply
that $\sin^2 (\alpha - \beta) < 0.1 $ if $M_h \lsim 50 \gev$.
More than this, a strikingly new feature is that
the $Z$ can now couple to a pair of nonidentical spin-0 objects,
leading to new Higgs particle production mechanisms.
Of particular
interest\footnote{Note that there are no tree level $ZZA$ or $W^+ W^- A$
                  vertices in this theory.}
is the process $Z^\ast \ra h + A$,
with a rate $\propto \cos^2(\alpha-\beta)$. A lack of such events at
LEP1 can then be translated to a constraint in the three-dimensional
($M_h, M_A, \alpha - \beta$) parameter space~\cite{SM_hig_lep}.
For a given ($M_h, M_A$) combine, this obviously translates to an
relatively strong upper limit on $\cos^2 (\alpha - \beta)$.
The two constraints are thus complementary to each other.
In addition, one has also to consider the fact that a non-trivial
Higgs sector may lead to additional contribution to the $Z$
width, even if the new decay channels cannot be identified over the
SM background. Yet, a light Higgs pair
($M_h + M_A \lsim 70 \gev$) may still be accomodated~\cite{SM_hig_lep}.

In this Letter, we concentrate on the scenario wherein
either $h$ or $A$ is very light~\cite{gun_hab}. While,
for a light $h$, this clearly warrants that
$\alpha \simeq \beta$, it is not necessary if only $A$ is
light and $M_h > m_Z - M_A$. However, in order to reduce
the number of parameters and thus simplify the analysis,
we shall not only impose this constraint, but rather promote
it to an exact equality. Since we propose to use
charged lepton decay modes as our signal, we further restrict ourselves
to the scenario with a large $\tan \beta$ ($\gsim 20$).
Note that, for $M_{h,A} < 10 \gev $, this parameter is not
yet constrained by  LEP1 data. Although, for
$M_{h,A} \gsim 10 \gev $, non-observation of the Yukawa process
($Z \ra b \bar{b} h/A$) constrains $\tan \beta$ to be
below 10(5) for scalar (pseudoscalar)~\cite{tb_Yuk}, the same
process is unlikely to be as efficient for lower Higgs boson
masses. Such an analysis is currently in progress though~\cite{new}.
Low energy data like those on the muon anomalous magnetic moment
still allow $\tan \beta \sim 20$ or higher for
$M_h \gsim$ 2--3~ GeV~\protect\cite{early_theory,maria} (see later discussion).

\begin{figure}[ht]
\vskip 4.55in\relax\noindent\hskip -1.55in
             \relax{\includegraphics{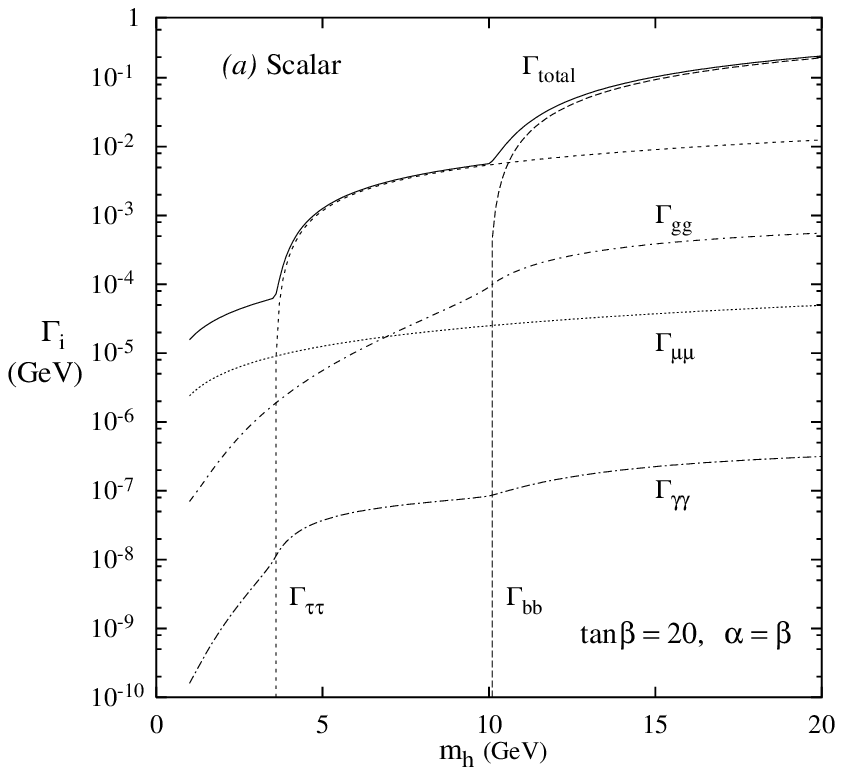}}
             \relax\noindent\hskip 3.35in
             \relax{\includegraphics{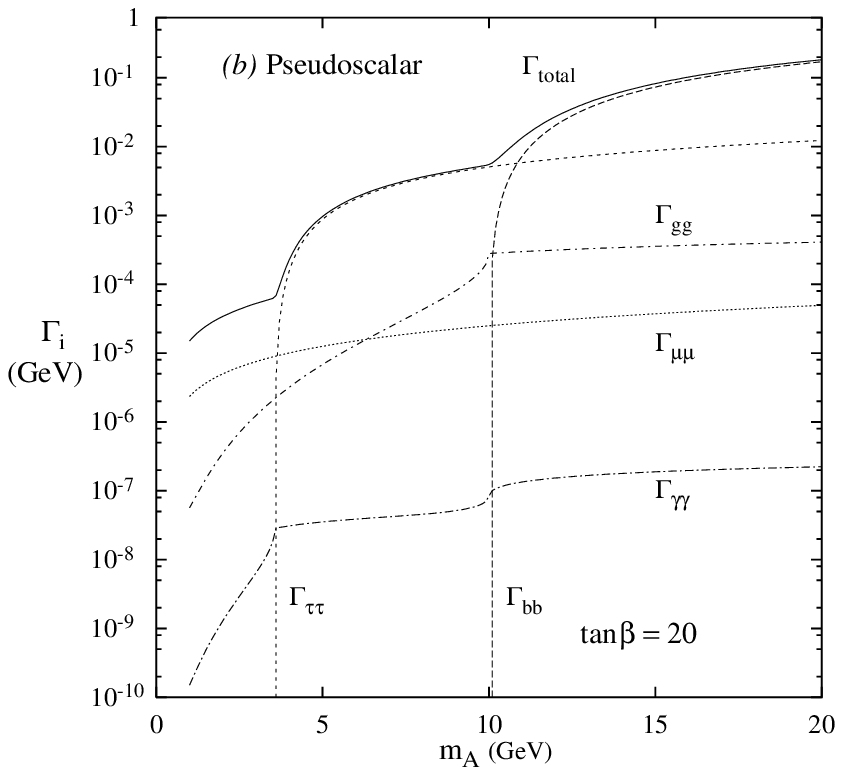}}
\vspace{-22.5ex}
\caption{ {\em The  partial and total
                    decay widths for $\tan \beta = 20$:
               {\em (a)} scalar $h$ ($ \alpha = \beta$), and
               {\em (b)} pseudoscalar $A$.} }
    \label{fig:width}
\end{figure}

In Fig.\ref{fig:width}, we present relevant for our analysis
  partial widths of $h$ and $A$
(for $\tan \beta = 20$) obtained under the above hypothesis.
To the leading order, all the widths shown in the
figure scale as $\tan^2 \beta$.
Note that the fermionic branching fractions for the two cases
follow a very similar pattern, and the only noticebale difference
occurs in the 2-photon and the 2-gluon decay modes.

Resonant neutral (pseudo)scalar production may occur at a
$\gamma \gamma$ collider through
two photon fusion at one loop. While only the charged fermion loops
contribute for $A$, the $\gamma \gamma h$ vertex would, in general,
receive corrections from $W^{\pm}$ and $H^{\pm}$ loops as well.
However, for $\alpha=\beta$ the $W^{\pm}$ contribution vanishes identically.
In the same limit, the $H^+ H^- h$ vertex is proportional to
$(g m_Z / 4 \cos\theta_W)\ \sin(4\beta)$, where $g$ is the weak coupling
constant. Clearly, this vertex becomes progressively weaker as $\tan \beta$
increases beyond 20. Moreover, this contribution weakens further as
$M_{H^+}$ increases. Since we assume that the charged scalars are indeed
heavy, this contribution can be safely neglected for the purpose
of our study.

The cross section for the  basic process
$\gamma \gamma \ra h \ra f {\bar f}$ (where we specify $f$ to be
$\tau$ or $\mu$ as the most important decay modes) is given
by\footnote{A similar expression holds for the pseudoscalar.}
\be
\sigma_{\gamma \gamma} =
  {{8\pi \Gamma (h\ra \gamma \gamma) \Gamma(h\ra f {\bar f})}
  \over{ (s_{\gamma \gamma} -M_h^2)^2+\Gamma^2_h M_h^2}}
  \; (1 + \lambda_1 \lambda_2) \ ,
    \label{prodn}
\ee
where $\lambda_i$ are the mean helicities of the photon beams
and the rest of the symbols carry their usual meaning.
In order to calculate the total cross section $\sigma^h_{ee}(f \bar{f})$,
we need to fold the above with the appropriate photon spectrum.
Thus, for an $e^+ e^-$ center of mass of $\sqrt{s_{ee}}$, the
differential cross section is given by
\be
\frac{{\rm d} \sigma^h_{ee}}{ {\rm d} s_{\gamma \gamma} }
    = \int_{x_{\rm min}}^{x_{\rm max}} \frac{ {\rm d} x_1 } {x_1 s_{ee}} \;
                 \sigma_{\gamma \gamma} \;
                                 f(x_1) \;
                     f\left( \frac{ s_{\gamma \gamma} } {s_{ee} x_1}
                          \right),
\ee
where $x_1$ (and $ x_2 \equiv s_{\gamma \gamma}/ s_{ee} x_1$) are the
momentum fractions of the initial electrons carried by the photons, and
${x_{\rm min}} = s_{\gamma \gamma}/ s_{ee}/ {x_{\rm max}}$.
The photon spectrum $f(x)$,  resulting
from Compton backscattering electrons on an intense laser light,
depends~\cite{telnov}
on the helicity of the initial electrons $\lambda_e$, initial
laser beam circular polarization $P_c$  and a machine
parameter $z$ that determines the maximum momentum carried by
the photon ($x_{\rm max} = z/(1 + z)$). Defining, for convenience,
variables $r \equiv x/(1-x)$ and $y \equiv 1 - x + 1/(1-x)$, we have,
for the photon spectrum and the mean helicity $\lambda_\gamma(x)$,
\be
\barr{rcl}
\displaystyle
{ {\rm d} n(x) \over {\cal N} {\rm d} x} & = & \displaystyle
y + 4 \frac{r }{z} \left( \frac{r }{ z } - 1 \right)
 -  \displaystyle { \lambda_e\: P_c\:} r (2 - x)
              \left( \frac{ 2 r }{z } - 1 \right)
\ , \\[3ex]
{ \lambda_\gamma (x)} & = & \displaystyle
\left( {{\rm d} n(x) \over {\cal N} {\rm d} x}  \right)^{-1}
\left[
- {  \lambda_e} r
   \left\{ 1 + (1 - x)
              \left( \frac{ 2 r }{z } - 1 \right)^2
   \right\}
  + { P_c}\: y \left( \frac{ 2 r }{ z } - 1 \right)
\right]\ ,
\earr
\ee
where ${\cal N}$ gives the normalization.

\begin{figure}[h]
\vskip 4.45in\relax\noindent\hskip -1.55in
             \relax{\includegraphics{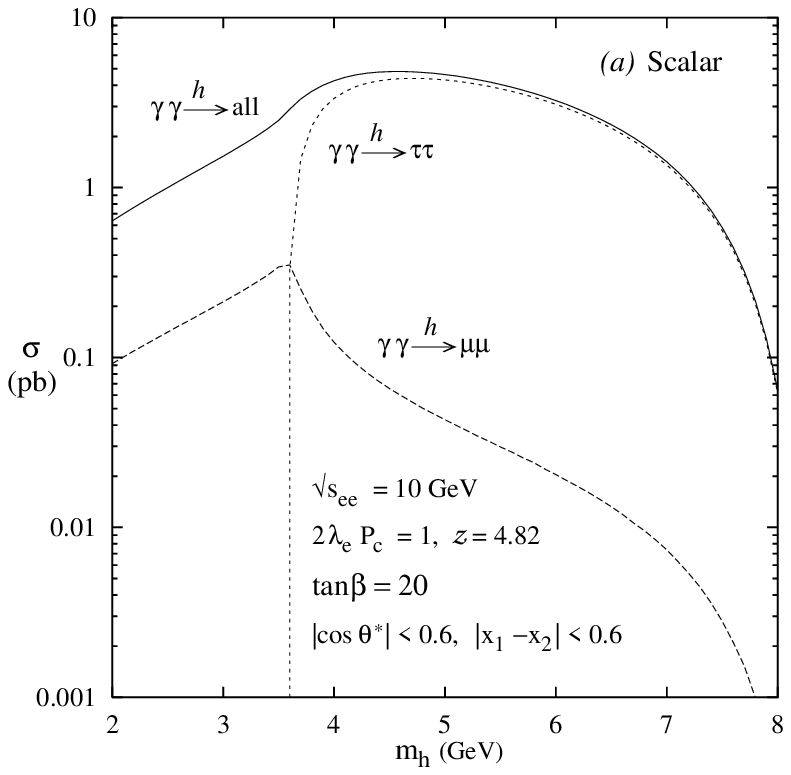}}
             \relax\noindent\hskip 3.35in
             \relax{\includegraphics{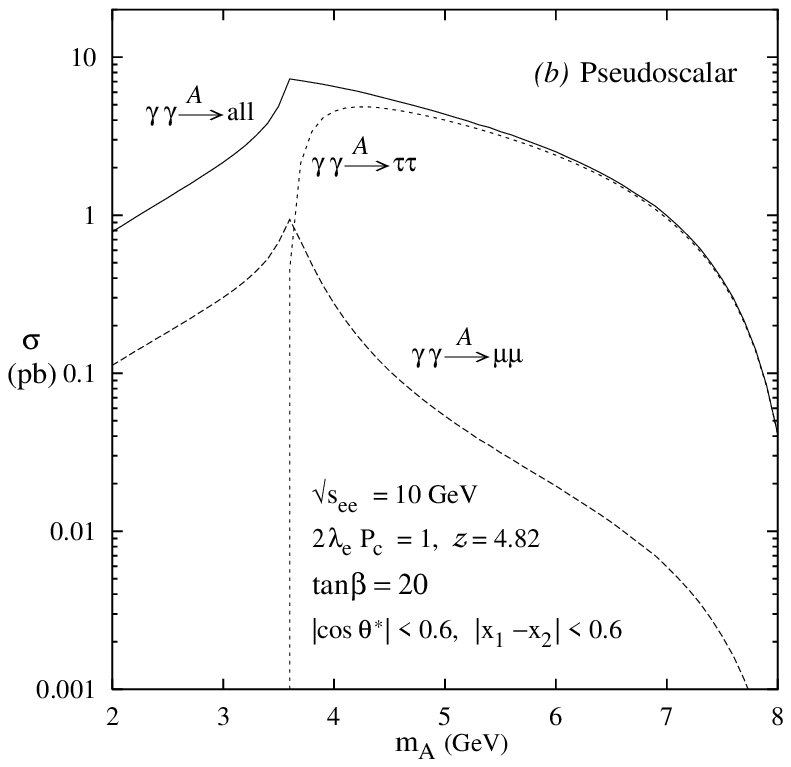}}
\vspace{-22.5ex}
\caption{ {\em The effective cross sections for the Higgs boson mediated
               process leading to $\mu^+ \mu^-$ and $\tau^+ \tau^-$
               final states ($\tan \beta = 20$). Also shown is the
               cross section summed over all decay channels.
               ({\em a}) scalar $h$  ($ \alpha = \beta$) and
               ({\em b}) pseudoscalar $A$.} }
    \label{fig:signal}
\end{figure}
We consider the resonant production of very light Higgs scalar
(pseudoscalar) at $e^+e^-$ NLC collider with energy
$\sqrt{s_{ee}}$=10 GeV~\cite{borden}. To maximize the photon energy,
and yet avoid multiple rescattering or pair--creation~\cite{telnov},
we choose $z = 2 (\sqrt{2} + 1) = 4.82$, and thus
$\sqrt{s_{\gamma \gamma}^{\rm max}} \simeq 0.83 \sqrt{s_{ee}} = 8.3 \gev$.
Following ref.~\cite{telnov}, we assume the `broad' spectrum
of photons with $2\lambda_e P_c=+1$. This has the advantage of
being rather flat over $\sqrt{s_{\gamma \gamma}}$ and, more importantly,
of favoring the $J_Z=0$ state,  the polarization state of Higgs scalar.
To a very good approximation,
$\sigma^h_{ee}(f {\bar f}) \propto \tan^2 \beta$.
In Fig.\ref{fig:signal}, we present the cross sections for
$\tan \beta = 20$. The cuts on the center of mass angle $\theta^*$
and the difference in the photon momentum fraction are motivated
below.

\begin{figure}[ht]
\vskip 4.45in\relax\noindent\hskip -0in
             \relax{\includegraphics{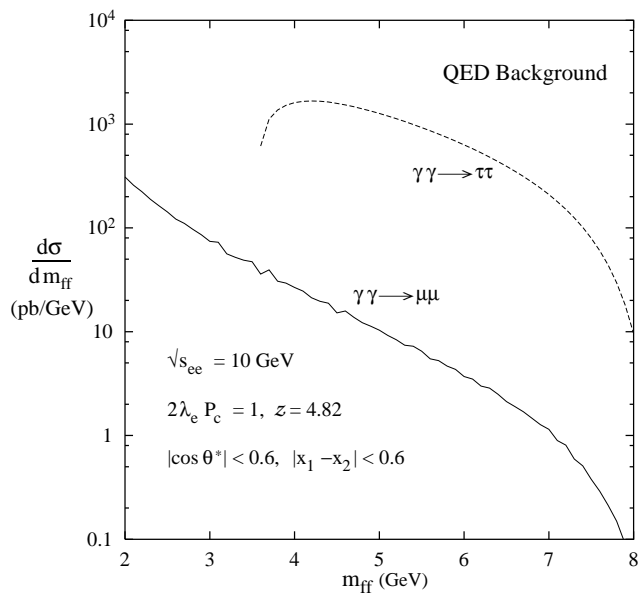}}
\vspace{-22.5ex}
\caption{ {\em The invariant mass distribution for the
               QED process $\gamma \gamma \ra f \bar{f}$.} }
    \label{fig:bkgd}
\end{figure}

The (orders of magnitude larger) background
is mainly due to direct $f {\bar f}$
pair production, since at such low energy the resolved
photon contributions are negligible. In Fig.\ref{fig:bkgd} we
show the invariant mass distribution for the QED process.
While the predominant
contribution ($J_Z = 2$) is already reduced significantly
by the choice for the spectrum, the forward/backward peaked $J = 0$
contribution is reduced by imposing a cut ($ -0.6 < \cos \theta^* < 0.6$)
on the CM scattering angle. As the signal is independent of
$\theta^*$, this cut eliminates only 40\% of the Higgs events.
It should be noted that this cut is more
effective for the muonic channel than for the tauonic channel on
account of the smaller mass of the muon. Although similar
reduction can be obtained for the $\gamma \gamma \ra \tau^+ \tau^-$
by further restricting $\theta^*$, this does not lead to an
appreciable improvement in the signal to background ratio.
Further improvements in this ratio can be made by restricting the total
boost of the $f \bar{f}$ system, or equivalently, by restricting
the difference in the momentum fractions
carried by the two colliding photons. We find that $|x_1 - x_2| < 0.6$
is an optimal choice.
\begin{figure}[ht]
\vskip 4.45in\relax\noindent\hskip -1.55in
             \relax{\includegraphics{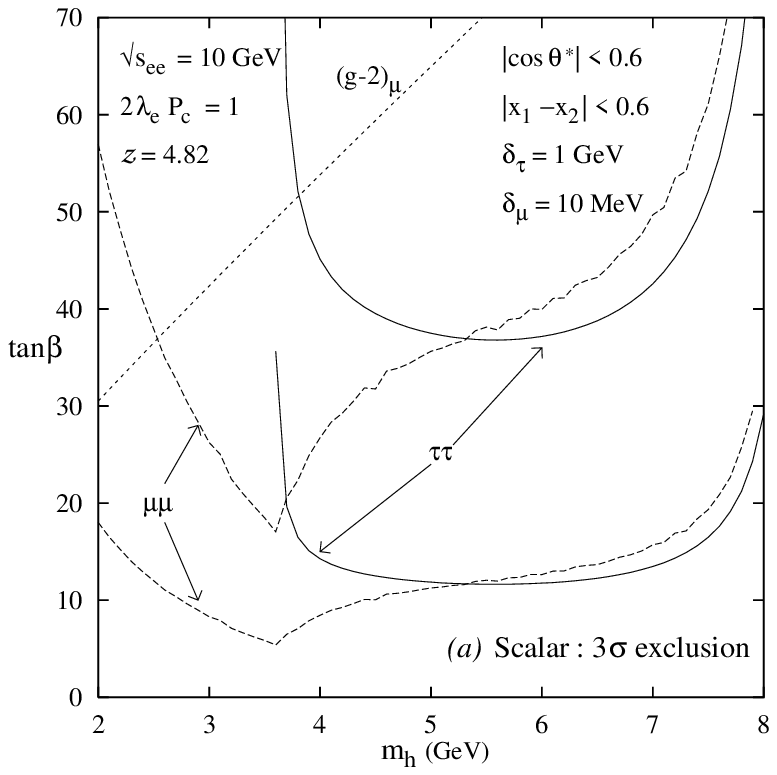}}
             \relax\noindent\hskip 3.35in
             \relax{\includegraphics{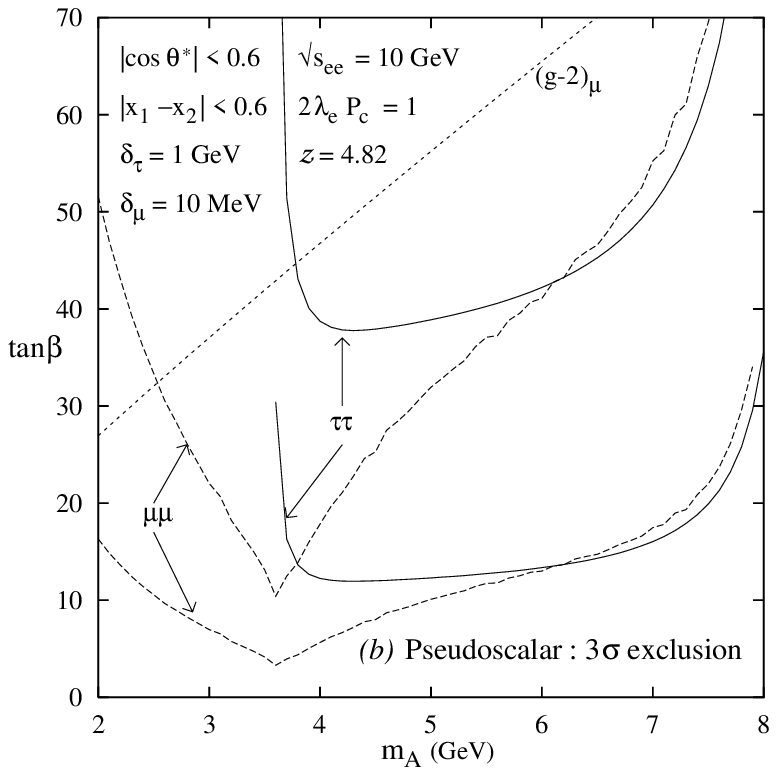}}
\vspace{-22.5ex}
\caption{ {\em The exclusion plots in the {\em (a)} $M_h$--$\tan \beta$
      and {\em (b)} $M_A$--$\tan \beta$ planes
      that may be achieved at NLC$\gamma\gamma$ from
      either of $\mu \mu$-- and $\tau \tau$--channels.
      Parameter space above the curves can be ruled out at
      $3 \sigma$ (99.7\% C.L.). The upper and lower sets are for
      integrated luminosity of 100 pb$^{-1}$ and 10 fb$^{-1}$ respectively.
      Also shown are the limits from the current data on $(g - 2)_\mu$
      under the assumption that the only non-SM contributions
      accrue from a light $h/A$.} }
    \label{fig:excl}
\end{figure}

It is clear from eqn.(\ref{prodn}), that the signal would have a
a sharp peak in the $f \bar{f}$ invariant mass. The task then is
to look for it over the continuous, but much larger, QED background.
Since $\Gamma_h$ ($\Gamma_A$) is tiny (see Fig.~\ref{fig:width}),
in the event of infinite resolution in the invariant mass, the
signal would be striking indeed! We adopt, though, a more realistic
approach and smear the signal (as well as the background) profile
with a gaussian resolution function~\cite{bbb}. Thus,
\be
\frac{{\rm d} \sigma^h_{ee} }{ {\rm d} m_{f \bar{f} } }
    = \frac{1}{\delta_f \sqrt{2\pi} }
      \int_{4 m_f^2}^{x^2_{\rm max} s_{ee}} {\rm d} s_{\gamma \gamma} \;
           \frac{{\rm d} \sigma}{ {\rm d} s_{\gamma \gamma} } \;
           \exp \left(- \frac{ (m_{f \bar{f}}
                                - \sqrt{\sigma_{\gamma \gamma} } )^2}
                             { 2 \delta_f^2} \
                \right).
\ee
For the experimental resolutions, we choose
$\delta m_{f {\bar f}} = 2 \delta_f$, with
 $\delta_{\mu }$=0.01 GeV and $\delta_{\tau}$=2 GeV.
The $3 \sigma$ (99.7\% C.L.) exclusion plots in the
$\tan \beta $--$M_{h/A}$ plane that may then be
 obtained using the $\mu^+ \mu^-$ and the $\tau^+ \tau^-$
final states are displayed in Fig.\ref{fig:excl}.
To be specific, we have adopted two representaive values
for the integrated luminosity : 100 pb$^{-1}$ and
 10 fb$^{-1}$. It is interesting that though the
$\tau^+ \tau^-$ cross sections are typically larger, yet better
limits are obtained from the muonic channel. The reasons are twofold :
{\em (i)} as we have noted earlier, the angular cuts are more effective
in eliminating the QED muons than the tau's and {\em (ii)} the
invariant mass resolution for a $\tau^+ \tau^-$ pair is expected to be
much worse than that for the $\mu^+ \mu^-$ pair. In the ideal case
where $\delta_{\tau} \simeq \delta_{\mu}$, the bounds from the
two channels would be similar. Note that above results are not expected
to have large theoretical uncertainties in contrast to
\eg, the process~\cite{early_expt,maria} $\Upsilon \ra \gamma h (A)$
which, in principle, is sensitive to the same mass range.

Also displayed in Fig.\ref{fig:excl} are the bounds that can be obtained
from the measurement~\cite{pdg} of the anomalous magnetic moment of the
muon under the assumption that a light Higgs particle constitutes the only new
source of contribution.
Although the SM central value for this quantity depends on the
evaluation of hadronic vacuum polarization, the difference is miniscule.
For the curves above, we use Case A of ref.~\cite{nath} as this provides
the stricter bounds. A look at the figures tell us then that,
even for the low luminosity version of NLC$\gamma \gamma$, the
bounds obtainable from the experiment suggested here
would be much stronger than the current ones.
It should be borne in mind though that a substantial improvement in the
experimental measurement of $(g -2)_\mu$ is in the offing~\cite{brookhaven}
and if, in addition, theoretical errors could be reduced, this measurement
could provide much stronger constraints.

To summarize, we consider
the physics potential of high luminosity, low energy NLC$\gamma\gamma$
in the search for  a possible light Higgs (pseudo)scalar in the general
2HDM. The limits obtainable with an integrated luminosity of
100 pb$^{-1}$ would already be better for the mass range
$3 \gev \lsim M_h \lsim 8 \gev$ than the
existing limits from  $(g-2)_\mu$. A luminosity of the
order 10 fb$^{-1}$/y would of course lead to much more stringent bounds.
The latter compare very well with
those expected from a  20-fold improvement in $(g-2)_{\mu}$
measurement and do not depend upon any additional assumptions
regarding the spectrum of new physics. Photon polarization is crucial though.

{\bf Acknowledgements}:
MK would like to thank G.~Belanger and F.~Boudjema
for the important suggestion which led to these
calculations. DC would like to thanl M.~Baillargeon for some
illuminating correspondence.

\end{document}